\documentclass[]{article}

\usepackage[sort&compress,numbers]{natbib}
\usepackage{amsmath, amssymb, amsfonts, amsthm}
\usepackage{xcolor}
\usepackage[font=small,labelfont=bf,tableposition=top]{caption}
\usepackage[caption=false]{subfig}
\usepackage{ulem}
\usepackage{enumitem}
\usepackage{pgf}
\usepackage{lineno,hyperref}
\usepackage{cleveref}
\modulolinenumbers[5]

\newtheorem{defi}{Definition}

\crefformat{equation}{\textnormal{Eq. #2#1#3}}
\crefrangeformat{equation}{\textnormal{Eqs. #3#1#4 - #5#2#6}}
\crefmultiformat{equation}{\textnormal{Eqs. #2#1#3}}
{ and #2#1#3}{, #2#1#3}{ and #2#1#3}
\crefformat{figure}{Fig. #2#1#3}
\crefrangeformat{figure}{Figs. #3#1#4 - #5#2#6}
\crefmultiformat{figure}{Figs. #2#1#3}
{ and #2#1#3}{, #2#1#3}{ and #2#1#3}
\crefformat{table}{Table #2#1#3}
\crefrangeformat{table}{Tables #3#1#4 - #5#2#6}
\crefmultiformat{table}{Tables #2#1#3}
{ and #2#1#3}{, #2#1#3}{ and #2#1#3}
\crefformat{appendix}{Appendix #2#1#3}
\crefrangeformat{appendix}{Appendixes #3#1#4 - #5#2#6}
\crefmultiformat{appendix}{Appendixes #2#1#3}
{ and #2#1#3}{, #2#1#3}{ and #2#1#3}
\crefformat{section}{Section #2#1#3}
\crefrangeformat{section}{Sections #3#1#4 - #5#2#6}
\crefmultiformat{section}{Sections #2#1#3}
{ and #2#1#3}{, #2#1#3}{ and #2#1#3}
\crefformat{defi}{Definition #2#1#3}
\crefrangeformat{defi}{Definitions #3#1#4 - #5#2#6}
\crefmultiformat{defi}{Definitions #2#1#3}
{ and #2#1#3}{, #2#1#3}{ and #2#1#3}
\crefformat{thm}{Theorem #2#1#3}
\crefrangeformat{thm}{Theorems #3#1#4 - #5#2#6}
\crefmultiformat{thm}{Theorems #2#1#3}
{ and #2#1#3}{, #2#1#3}{ and #2#1#3}
\crefformat{lema}{Lemma #2#1#3}
\crefrangeformat{lema}{Lemmas #3#1#4 - #5#2#6}
\crefmultiformat{lema}{Lemmas #2#1#3}
{ and #2#1#3}{, #2#1#3}{ and #2#1#3}
\crefformat{corolario}{Corollary #2#1#3}
\crefrangeformat{corolario}{Corollaries #3#1#4 - #5#2#6}
\crefmultiformat{corolario}{Corollaries #2#1#3}
{ and #2#1#3}{, #2#1#3}{ and #2#1#3}

\begin{document}

	\begin{center}{Measuring COVID-19 spreading speed through the mean time between infections indicator}
		\end{center}
	
		\begin{center}		{Gabriel Pena}, {Ver\'onica Moreno}, and 	{N\'estor Ruben Barraza}\\
		{Universidad Nacional de Tres de Febrero}
		
	\end{center}

	
	
	
	\begin{abstract}
		{\bf Objectives:} To introduce a novel way of measuring the spreading speed of an epidemic. 
		\newline
		{\bf Methods:} We propose to use the mean time between infections (MTBI) metric as obtained from a recently introduced non-homogeneous Markov stochastic model. Different types of parameter calibration are performed. We estimate the MTBI using data from different time windows and from the whole stage history and compare the results. In order to detect waves and stages in the input data, a preprocessing filtering technique is applied.
		\newline
		{\bf Results:} The results of applying this indicator to the COVID-19 reported data of infections from Argentina, Germany and the United States are shown. We find that the MTBI behaves similarly with respect to the different data inputs, whereas the model parameters completely change their behaviour. Evolution over time of the parameters and the MTBI indicator is also shown.
		\newline
		{\bf Conclusions:} We show evidence to support the claim that the MTBI is a rather good indicator in order to measure the spreading speed of an epidemic, having similar values whatever the input data size. 
	\end{abstract}
	

	\section*{Introduction} 
	
	The COVID-19 outbreak, declared pandemic in 2020, attracted the attention of scientists from different domains (biologists, physicists, engineers and mathematicians, among others). The urgent need to control, predict and monitor the disease progress made it essential to count with mathematical models and algorithms (see for example \cite{cori2013new}), both to manage data on infections and deaths and to perform calculations and predictions. Not only were the well known SIR (Susceptible - Infectious - Removed) compartmental model and its derivations vastly applied \cite{kermack,SIMON20204252,CAO2020124628,liulevy,ali2020analysis,rojas2020comment,royal_sir,gleeson2022calibrating} but also many new models \cite{em_growth_gompertz} and AI algorithms \cite{DASILVA,SILVA,epidemics_likelihood_logistic,fokas2021covid} were proposed. One is the stochastic model presented in \cite{Chaos2020BPM} and \cite{MACI2021_MorenoPenaBarraza}, which we call the BPM model. It is based on a contagion Markov model described by a non-homogeneous birth process (NHBP). The authors were able to obtain the functional form of the cumulative infection cases and deaths curves, which have a sub-exponential shape, which is a behaviour previously pointed out for epidemics \cite{CHOWELL201666,VIBOUD201627,GANYANI2020110029,epidemics_logistic}. The model has two parameters: one that represents the power of the outbreak and another that models the immunization rate. The expected time between events can be obtained as a standard calculation in counting stochastic processes \cite{BookChapter2022}. Thus, useful indicator to measure the outbreak speed is obtained: the mean time between infections (MTBI). This indicator expresses the spreading speed, since less time between successive infections implies a more rapid disease spread. Therefore, it is expected that during the initial stage of a disease wave the MTBI indicator decreases as the peak gets closer. As we will show, taking the model parameters individually does not suffice to gain insight on the progress of the epidemic, but the MTBI actually does, which makes it a quite robust epidemiological indicator. Thus, the MTBI can be used to evaluate the impact of the actions taken by health institutions. 
	
	To calibrate the model parameters, we fit the data of the total reported cumulative infection cases to the mean value function of the process. Since this function does not have concavity changes, the data curve needs to be split into sections (which we will call stages) with a fixed concavity, and then the model can be separately applied to each of them. To achieve this, we look for the local maximums and minimums of the daily data curve, which cannot be directly observed due to sharp variations in the reported data. These high frequency jumps are commonly associated to noise in signal processing. Thus, we apply a filtering routine to smooth the daily data curve and then a standard maximum and minimum detection algorithm. Once the stages are properly separated, we fit the model for some specific stages in Argentina, Germany and the United States by the same method described in \cite{Chaos2020BPM} to obtain the MTBI indicator which depends on the model parameters. 
	
	An interesting result of the filtering algorithm is the comparison between the filtered curves of daily cases and daily deaths, which shows that both have rather similar shapes, with a short time delay in between. This behaviour was observed in the three considered countries.
	
	In \cite{Chaos2020BPM}, the parameter estimation at a given time is performed using the whole history of the corresponding stage. A natural question is whether it is correct to use the full history, or just some part of it. Different approaches to this question were recently presented in the context of the SIR model \cite{MACI2021_UNS,cordelli2020time}. In all of the proposed cases there exists some arbitrariness regarding the choice of the data used to calibrate the models. In this work, we perform a comparative study between the possibilities of using different data inputs, for example using the whole stage history or short time windows of a certain number of previous days. We show statistically that the MTBI does not change significantly, whereas the model parameters do. Evolution over time of the MTBI and those parameters is also shown. The most important result obtained is that the MTBI indicator is invariant with respect to the size of the data used for calibration, which lets us conclude that the MTBI is a robust indicator with respect to the data used to estimate it.
	
	\section*{Materials and methods}\label{sec_methods}
	
	\subsection*{The BPM model}\label{sec_modelo}
	
	In this work we consider a particular model based on a non-homogeneous birth process (NHBP) which was recently proposed \cite{Chaos2020BPM}. As it is known, NHBPs are a particular class of continuous time Markov processes \cite{stroockmarkov,klugman2013loss} that model the growth of a population where individuals can only be born. In a NHBP, the probability of having $r$ individuals in a population at a given time $t$, $P_r(t)$ is given by \cref{def_nhb} as the solution of a recursive system of ordinary differential equations \cite{Feller1}:
	
	\begin{defi}\label{def_nhb}
		Let $P_r (t), \; r \in \mathbb{N}_0$ denote the probability of having $r$ individuals at time $t$. Then, for an initial population of $0$ individuals we have $P_0(0) = 1$ and $P_r(0) = 0$ $\forall r > 0$, whereas for $t > 0$ the pmf $P_r(t)$ is defined recursively as the solution of each differential equation in the system of ODEs,
		\begin{align}
			\label{nhb_dif_eq}
			P'_r(t) &= -\lambda_r(t) P_r(t) + \lambda_{r - 1}(t) P_{r-1}(t), \quad r>0, \nonumber \\
			P'_0  (t)&= -\lambda_0(t)P_0(t),
		\end{align}
		
		where $\lambda_r (t)$ can depend on both $t$ and $ r $. The function $\lambda_r(t)$ is called event rate or intensity function.
	\end{defi}
	
	The model considered here can be obtained as a particular case of the more general models presented in \cite{konno2010exact,sendova,klugman2013loss}. This particular process is governed by the following event rate:
	\begin{equation}
		\label{proposedfr}
		\lambda_r(t) = \rho \frac{1 + \frac{\gamma}{\rho} \; r}{1 + \rho \; t},
	\end{equation}
	where $\gamma, \rho >0$ are the model parameters. $\gamma$ measures the power of the spreading disease and $\rho$ determines the immunization rate. 
	
	Let $M(t)$ be the mean number of individuals of the population (mvf) for the model. In \cite{Chaos2020BPM}, the authors prove that:
	\begin{equation}
		\label{mean_ours}
		M(t)=\frac{\rho}{\gamma} \left[  (1 + \rho t)^{\frac{\gamma}{\rho}} - 1 \right].
	\end{equation}
	
	Quite interesting observations arise from \cref{mean_ours}. The mvf is a power of $t$ with exponent $\gamma/\rho$; hence, the form of the function may be quite different depending on the value of this ratio. There are three possible scenarios:
	\begin{itemize}
		\item Case 1: $\frac{\gamma}{\rho} < 1$. The curve is concave.
		\item Case 2: $\frac{\gamma}{\rho} = 1$. \cref{mean_ours} reduces to that of the Polya-Lundberg process, the curve is a straight line.
		\item Case 3: $\frac{\gamma}{\rho} > 1$. The curve is convex.
	\end{itemize}
	A natural consequence of this is that this process can model a large variety of different scenarios, including those which were accurately described by the Polya-Lundberg process \cite{lundberg1964random}. As shown in \cite{Chaos2020BPM}, the $M(t)$ function fits quite well the cumulative number of either infections or deaths from the COVID-19 pandemic, which allows estimating the parameters $\gamma$ and $\rho$ using the reported data by standard fitting techniques like Least Squares.
	
	An important indicator useful to measure the spreading speed is obtained from the model: the expected elapsed time between the occurrence of an event and the next one, which corresponds to the Mean Time Between Infections (MTBI) or the Mean Time Between Deaths (MTBD) depending on whether the considered population consists of the infected cases or deaths, respectively. In this section we use the general notation MTBE (Mean Time Between Events). For this particular model, the formula 
	\begin{equation}\label{MTBI_prop_fr}
		MTBE(t)= \frac{1}{\rho} \; \frac{1 + \rho t}{(1 + \rho t)^{\frac{\gamma}{\rho}} - 1}
	\end{equation}
	gives the mean time between the event that occurred at time $t$ and the next \cite{Chaos2020BPM}.
	
	\subsection*{Filtering algorithm}
	
	Since the mvf curve does not have inflection points, the model needs to be applied separately in each stage having either positive or negative concavities, which correspond to the stages (early/mitigation) of each outbreak wave. This implies that a criterion for the wave and stage separation needs to be defined. We choose a rather classic approach:
	\begin{itemize}
		\item Local minimums in the daily data report represent time instants where a wave finishes and a new wave starts.
		\item Local maximums in the daily data report represent time instants where, within a single wave, the initial stage ends and the mitigation stage begins. 
	\end{itemize}
	
	The main difficulty regarding these definitions is that the measured data (daily cases and daily deaths) is extremely noisy. This implies that finding local extrema by direct observation is not straightforward. To overcome this problem we perform a filtering procedure to suppress the noise and smooth the curve, followed by a maximum and minimum detection algorithm. In this work we considered Argentina, Germany and the United States as our study cases, but the same results shown here can be achieved for any other dataset by properly setting the filter hyper-parameters.
	
\begin{figure}[ht]
		\centering
		\subfloat[\centering Argentina]{\label{arg_cases_spectrum}\includegraphics[width = 0.48\linewidth]{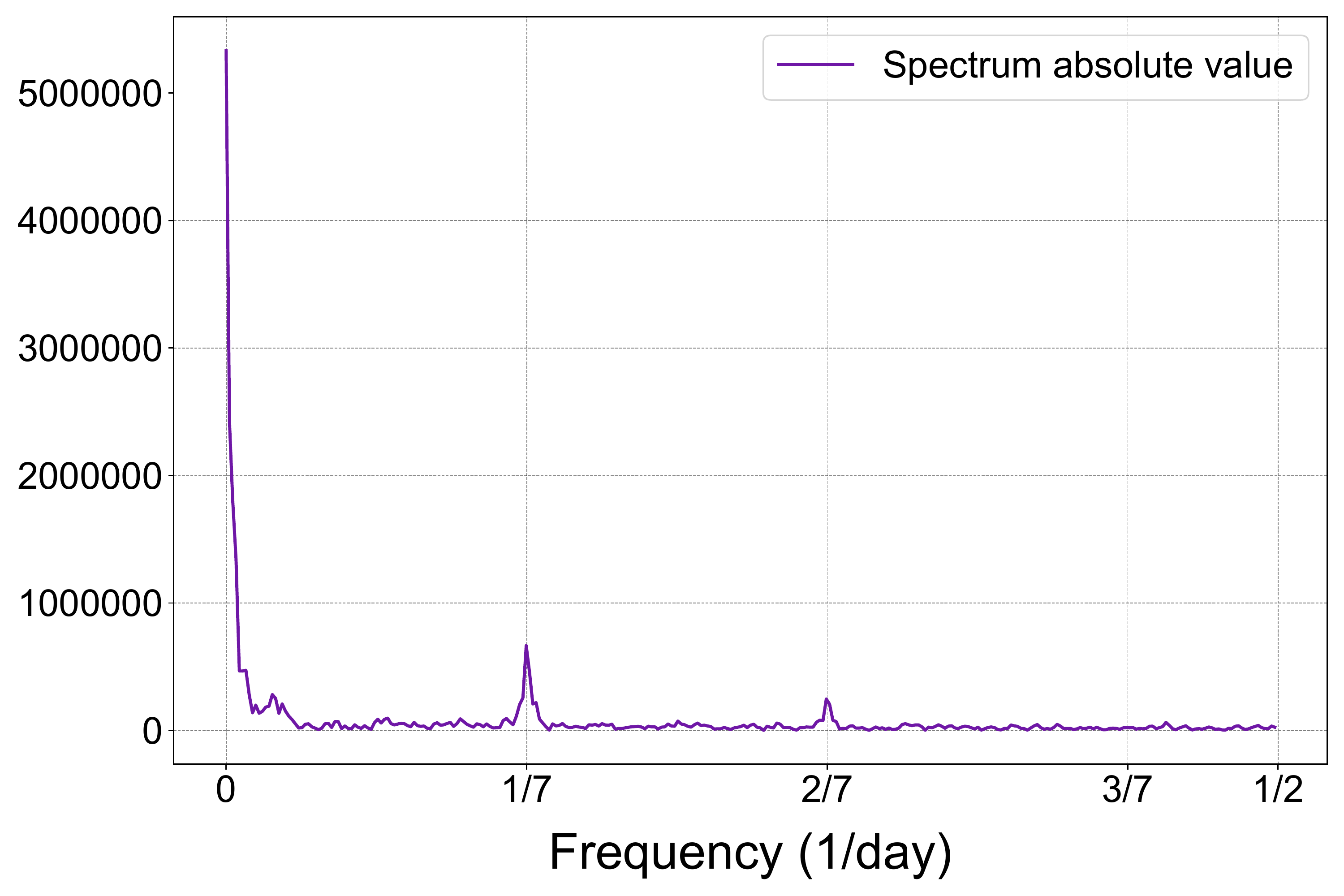}}
		\subfloat[\centering Germany]{\label{ger_cases_spectrum}\includegraphics[width = 0.48\linewidth]{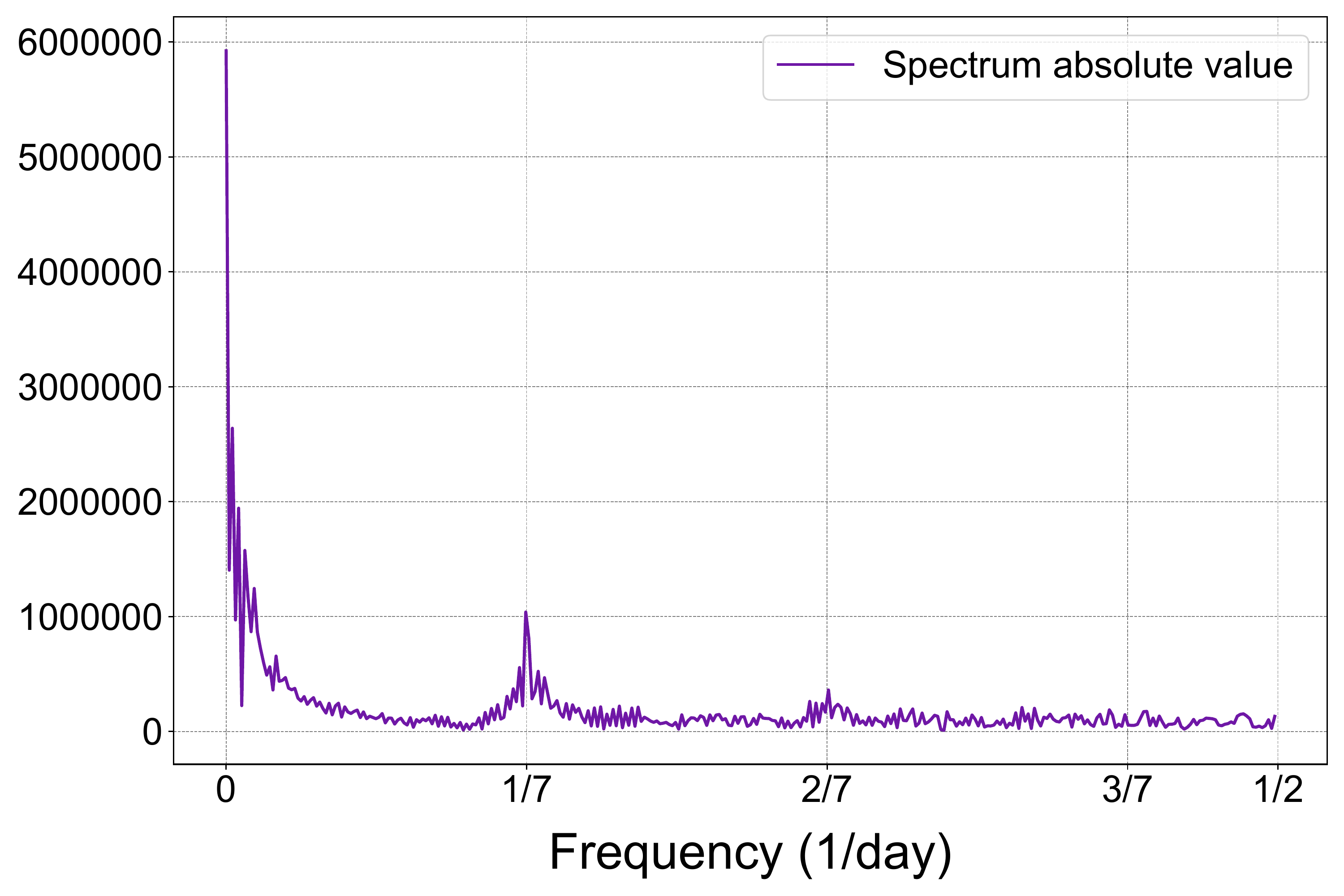}} \\
		\centering
		\subfloat[\centering United States]{\label{usa_cases_spectrum}\includegraphics[width = 0.48\linewidth]{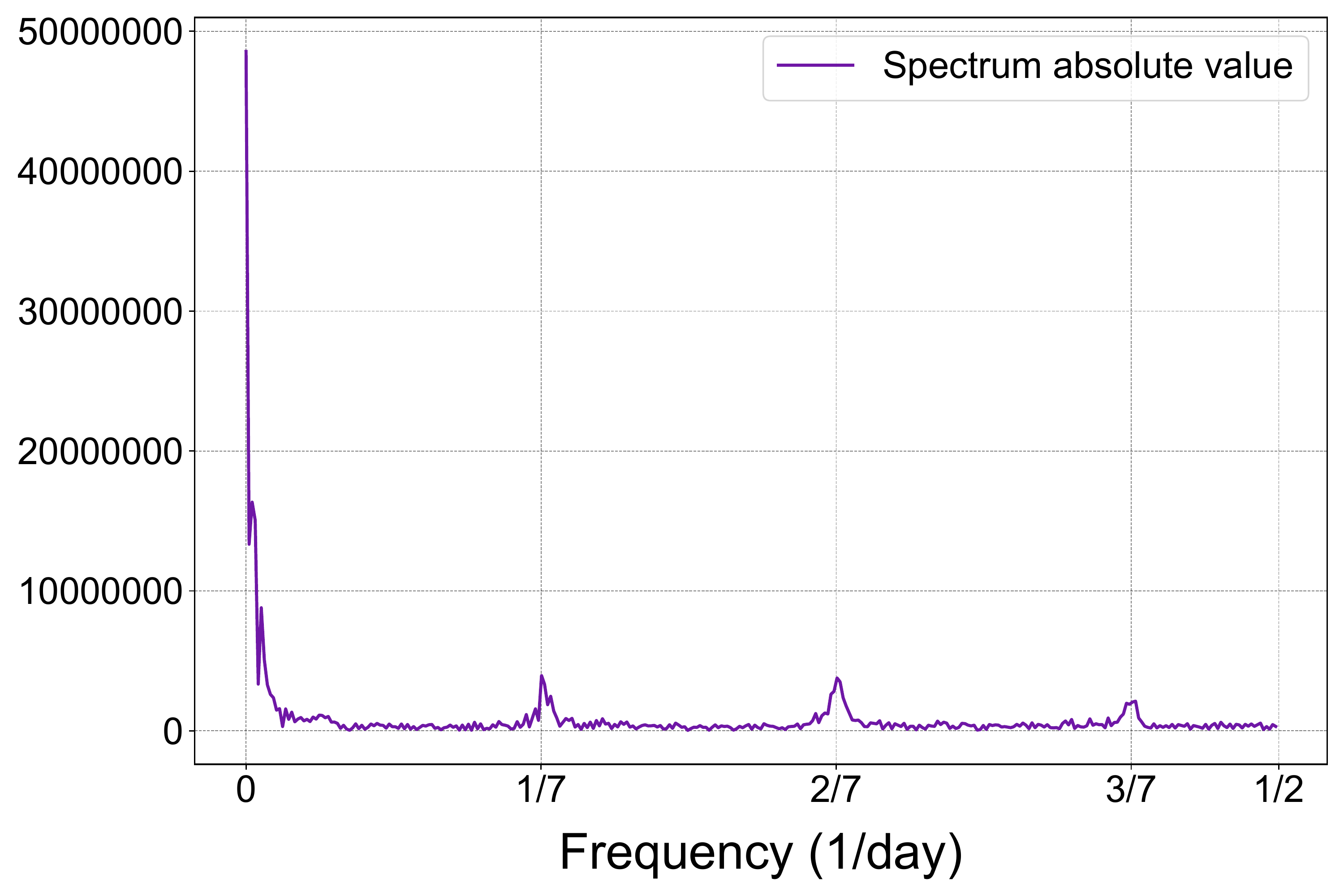}}
		
		\caption{Spectrum of the daily cases data, as obtained by applying the FFT algorithm. Frequency axis has $1/day$ units, and is shown up to 1/2, the highest observable frequency (due to the data being sampled at 1 datum per day).}
		\label{fig_spectrum}
	\end{figure}

	Some interesting observations regarding the noise can be made by looking at the data spectrum. \cref{fig_spectrum} depicts the absolute value of the Discrete Fourier Transform (DFT) of the daily reported cases in the three mentioned countries. Significant peaks can be clearly distinguished at frequency $1/7$ and its second harmonic (and its third in the United States case), which is explained by the ``once per week'' periodic variations in the testing amount on weekends, which are known to be lower. An analogous behaviour can be seen in the deaths dataset. These observations evidence the need of a preprocessing, namely a low-pass filtering, in order to suppress measurement noise. Consequently, we use moving-average (MA) filters, also known as sliding windows, which are the simplest low-pass FIR filters; see \cite{SMITH2003277} for technical details about MA filters. We proposed to apply $n$ MA filters of length $L$ in series; $L$ and $n$ are the algorithm hyper-parameters. 
	
	\begin{figure}[ht]
		\centering
		\subfloat[\centering Argentina]{\label{arg_daily_cases}\includegraphics[width = 0.48\linewidth]{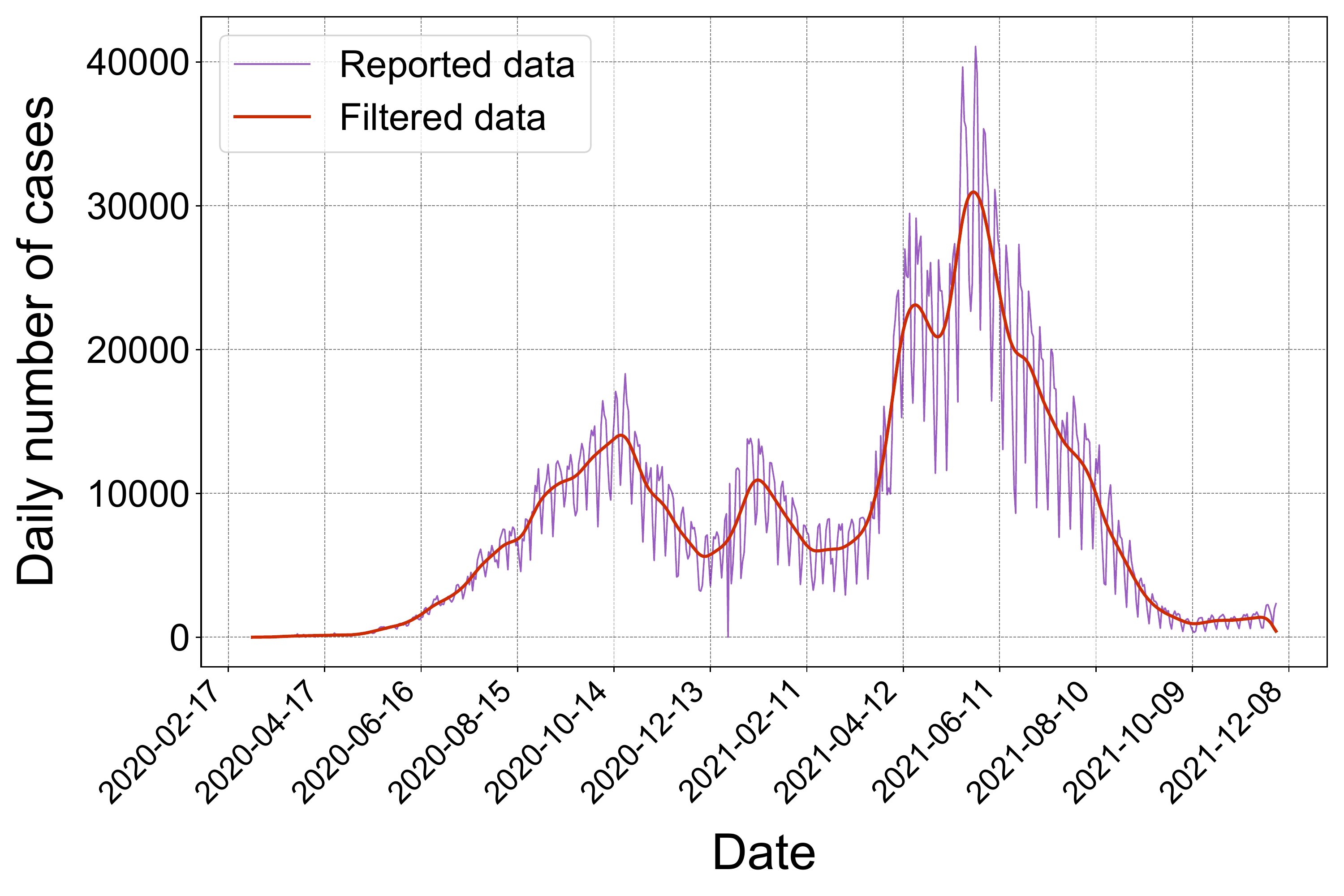}}
		\subfloat[\centering Germany]{\label{ger_daily_cases}\includegraphics[width = 0.48\linewidth]{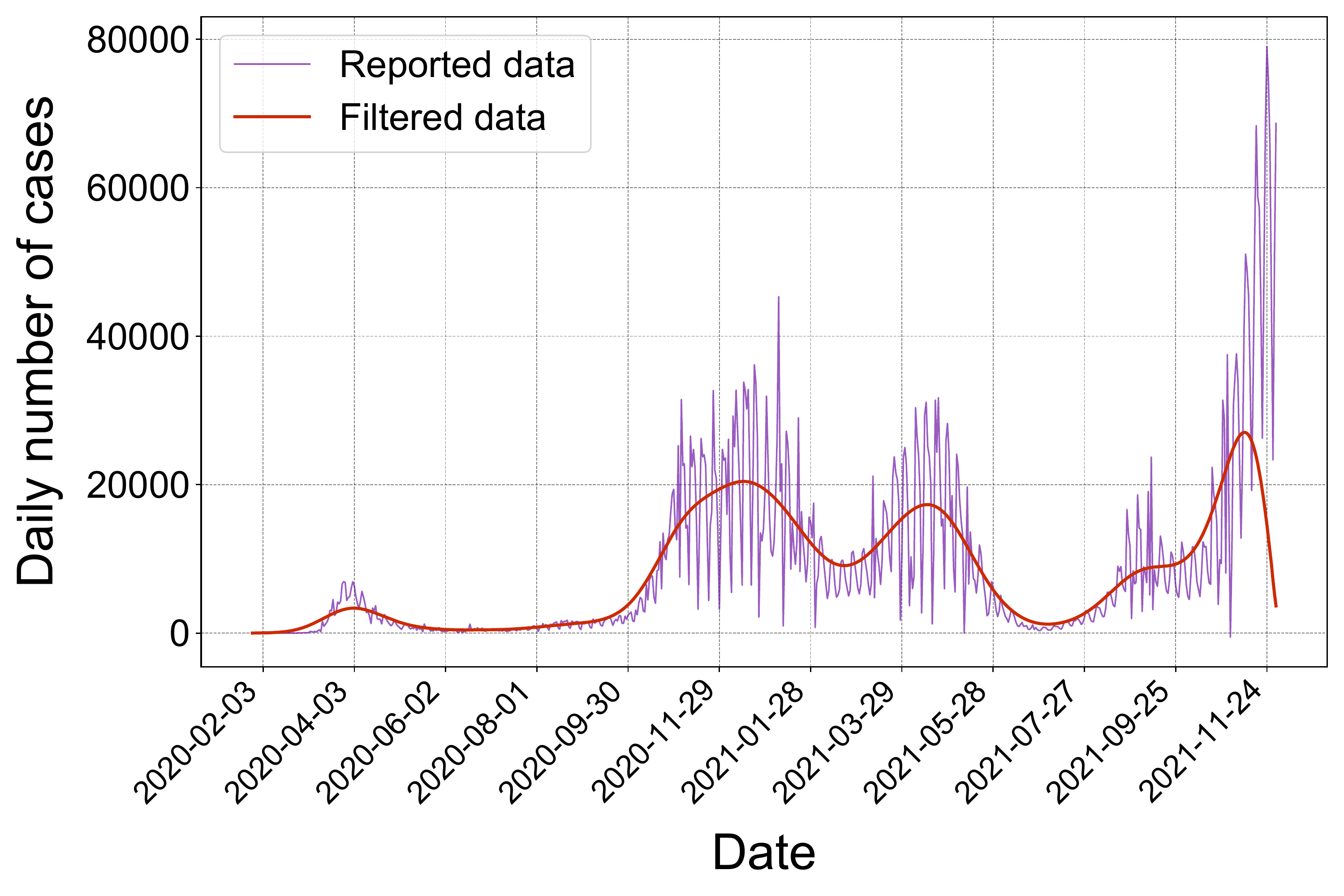}} \\
		\centering
		\subfloat[\centering United States]{\label{usa_daily_cases}\includegraphics[width = 0.48\linewidth]{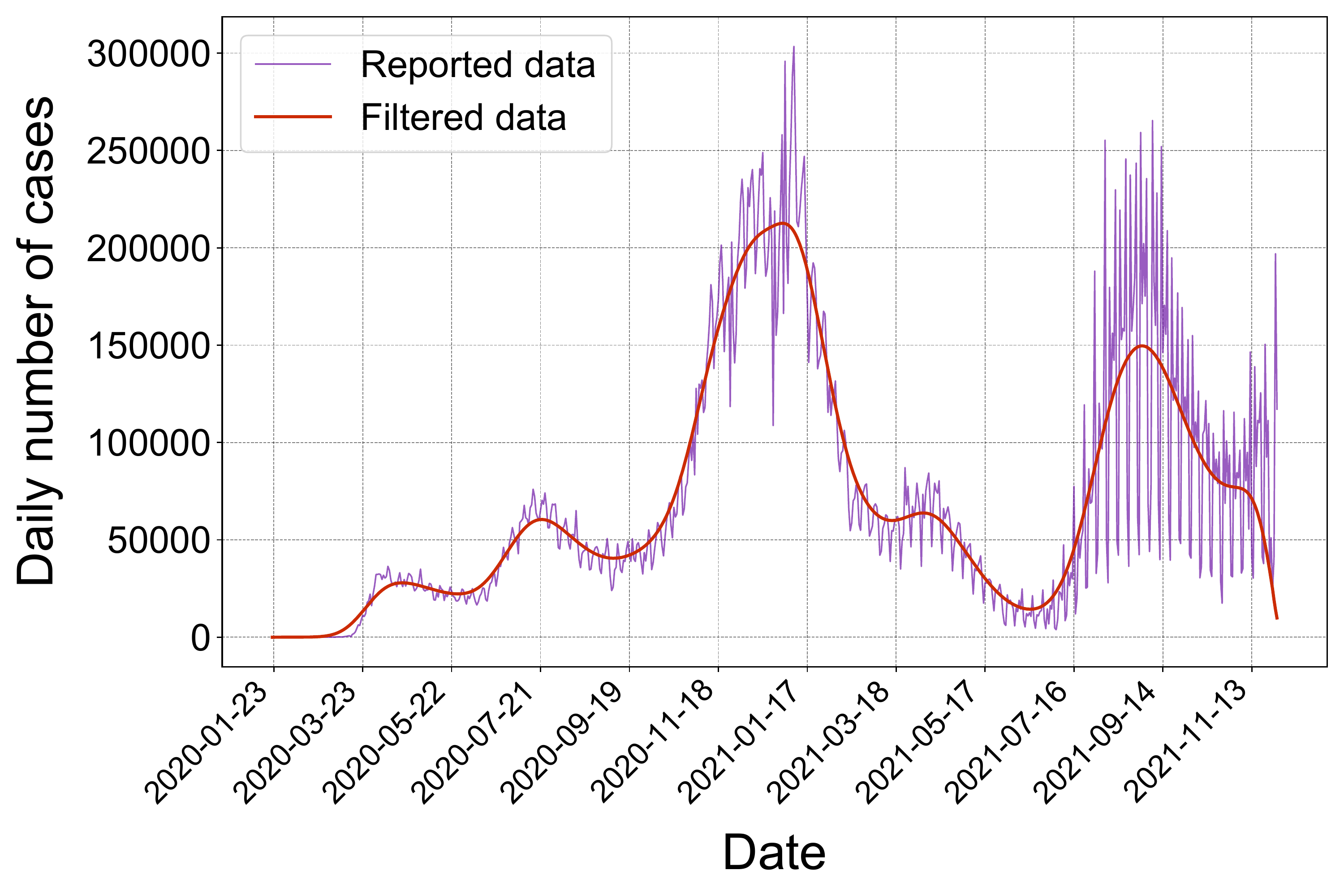}}
		
		\caption{Daily cases, reported (purple) and filtered (red).} 
		\label{fig_cases}
	\end{figure}
	
	The reported cases from Argentina, Germany and the United States and their respective filtered curves are shown in \cref{fig_cases}. Hyper-parameters were chosen not without some arbitrariness, and are not the same in the three cases. Since we want to remove the $1/7$ frequency component at least, $L$ must be chosen to be $\geq 7$. In order to avoid harming the signal too much, we chose to fix $L=7$ and increase $n$ as much as necessary, achieving a nearly Gaussian behaviour \cite{SMITH2003277} without suppressing useful signal frequencies. 
	
	Having a smooth curve, detecting maximums and minimums can be done by straightforward methods. It must be remarked that the algorithm is not completely automatic: proper hyper-parameter values must be carefully chosen by a human user, and since filtering is not perfect, in some cases human interpretation is necessary to determine which minimums and maximums are ``false'' and which are not. 
	
\begin{figure}[ht]
		\centering
		\subfloat[\centering Argentina]{\label{arg_filtered} \includegraphics[width = 0.48\linewidth]{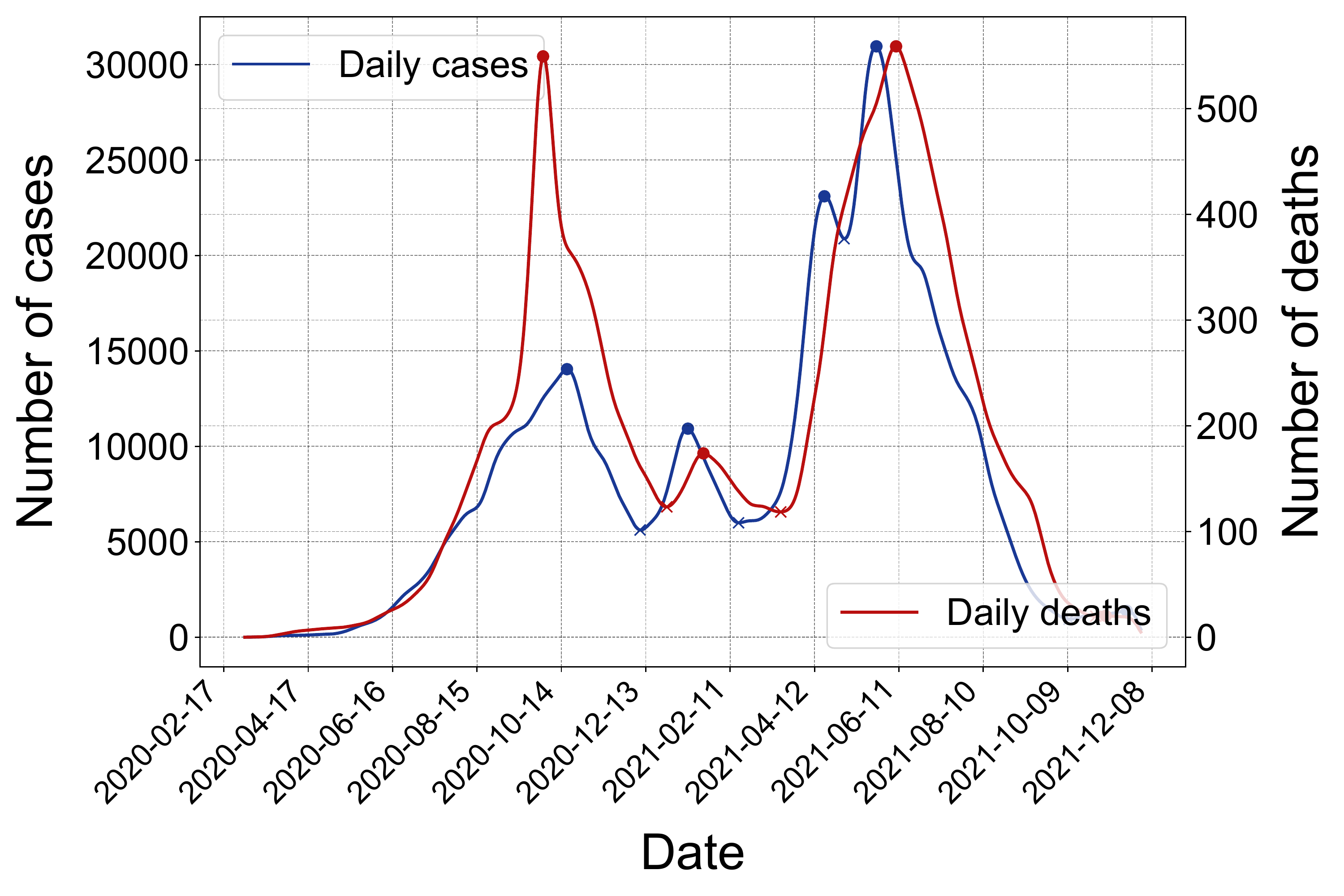}}
		\subfloat[\centering Germany]{\label{ger_filtered}\includegraphics[width = 0.48\linewidth]{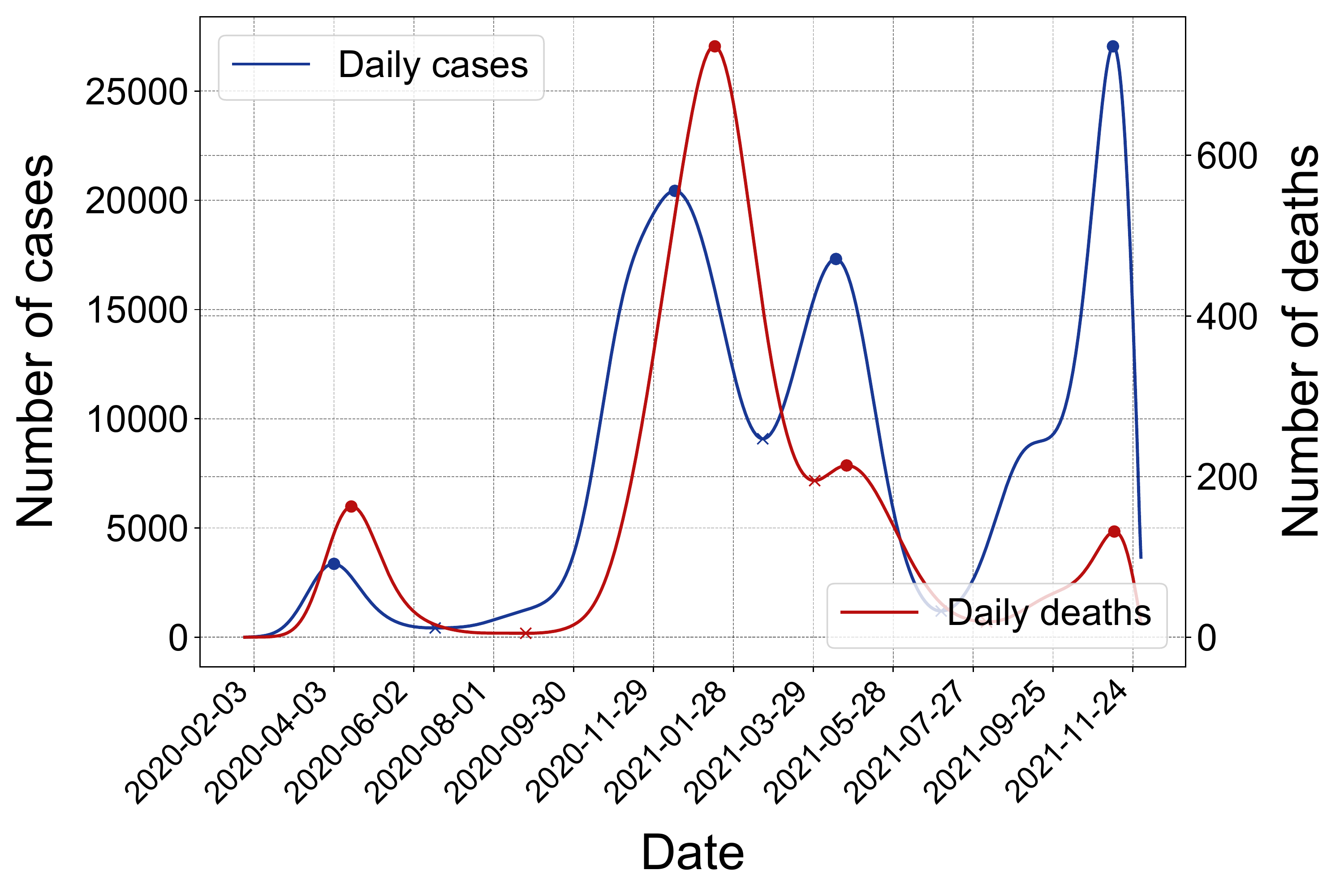}} \\
		\centering
		\subfloat[\centering United States]{\label{usa_filtered}\includegraphics[width = 0.48\linewidth]{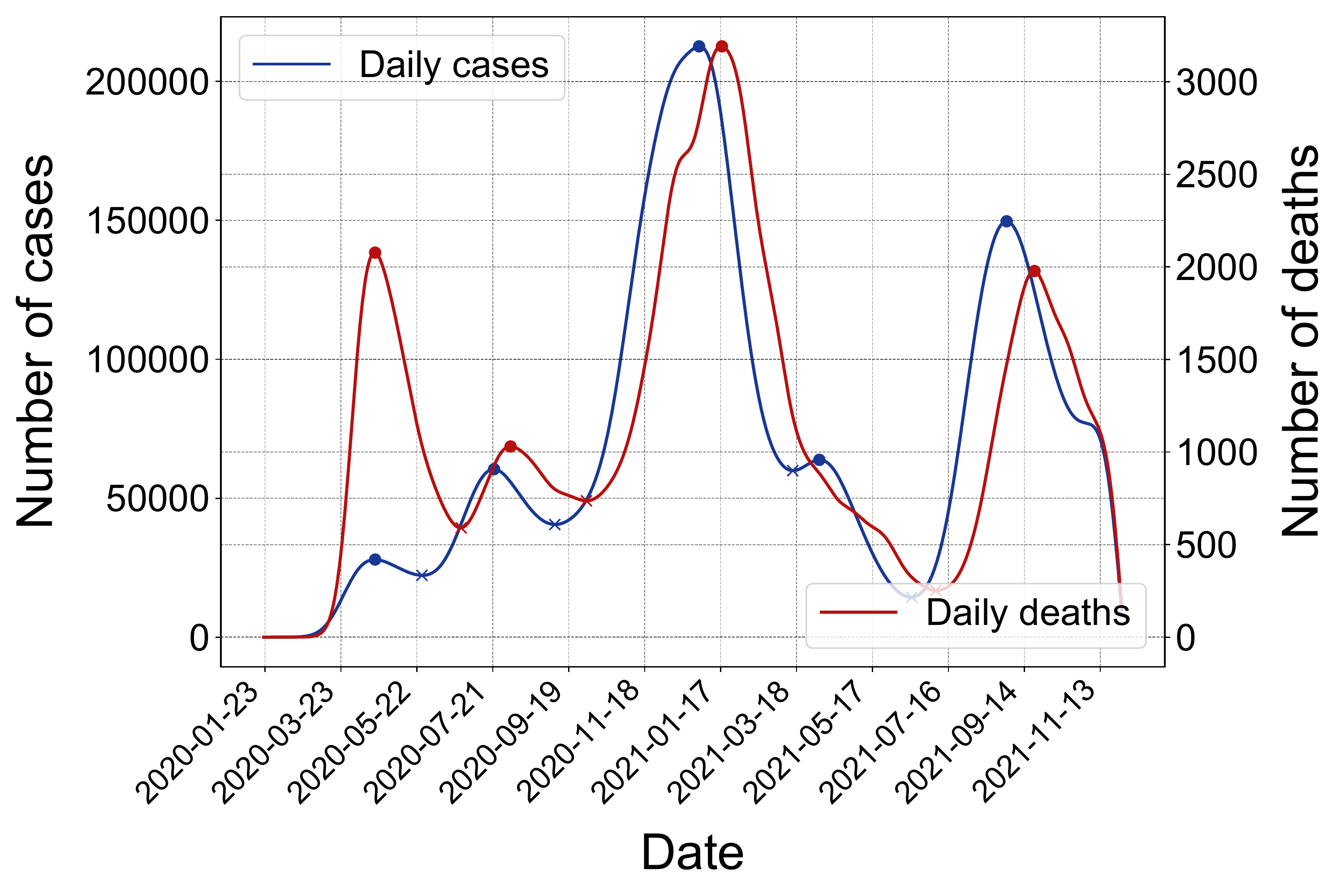}} 
		
		\caption{Daily cases (blue, scale shown on the left side axis) and deaths (red, scale shown on the right side axis) filtered curves. Filled circles indicate local maximums; $\times$ marks indicate local minimums.} 
		\label{fig_cases_deaths_output}
	\end{figure}

	In \cref{fig_cases_deaths_output} we show the filtered curves corresponding to the daily cases and daily deaths in a single plot in order to compare them. It is interesting to note that the cases and deaths curves have similar shapes, with the deaths curve delayed by a short period (approximately two weeks) as expected due to the length of the infectious period. This also shows that the propagation speed of deaths is directly related to the propagation speed of infections. However, due to the existence of measurement noise, they do not always exhibit the same amount of waves.
	
	\section*{Results} \label{sec_results}
	
	Once the data is properly segmented into sections, the propagation speed of the disease can be assessed by analyzing the evolution over time of the MTBI indicator within each of them. In order to do this, we are going to fit the model's mvf (\cref{mean_ours}) to the cumulative infection cases, recalibrating the parameters for each time $t$, obtaining this way two estimates $\hat{\gamma}(t)$ and $\hat{\rho}(t)$. Replacing these on \cref{MTBI_prop_fr} yields:
	\begin{equation}
		\label{mtbi_t}
		\widehat{MTBI}(t) = \frac{1}{\hat{\rho}(t)} \; \frac{1 + t \cdot \hat{\rho}(t)}{(1 + t \cdot \hat{\rho}(t))^{\frac{\hat{\gamma}(t)}{\hat{\rho}(t)}} - 1},
	\end{equation}
	which gives the evolution over time of the MTBI indicator estimate.
	
	The fittings were performed by the nonlinear Least-Squares method, using the Levenberg-Marquardt algorithm implemented in the Python language; the source code provided by the authors of \cite{Chaos2020BPM}, where they implemented the BPM model, is available at \cite{epydemics_source_repo}. The experiments were done using the real (unfiltered) data, since the cumulative curves do not change significantly after filtering. Datasets were obtained from \cite{owid_covid_data}.  
	
	A natural question arises when fitting the model: is the whole history of a stage significant, at any given time instant, or should a shorter time period be considered instead. This question has already been addressed in the context of the SIR model \cite{MACI2021_UNS, cordelli2020time}. We are going to perform a comparative study of the results obtained by calibrating the model parameters (for every time index $t$ within a stage) using the whole history of the stage up to $t$ and those obtained by considering different time windows prior to the respective $t$. As it is customary in signal processing, we call this procedure ``windowing''. For the analysis, we considered the following study cases:
	\begin{itemize}
		\item Argentina: initial stage of the first wave, from March 03, 2020 to October 17, 2020.
		\item Germany: initial stage of the fourth wave, from July 02, 2021 to November 08, 2021. 
		\item United States: initial stage of the third wave, from September 07, 2020 to December 30, 2020.
	\end{itemize} 
	
	\begin{figure}[ht]
		\centering
		\subfloat[\centering Argentina]{\label{arg_mtbi}\includegraphics[width = 0.33\linewidth]{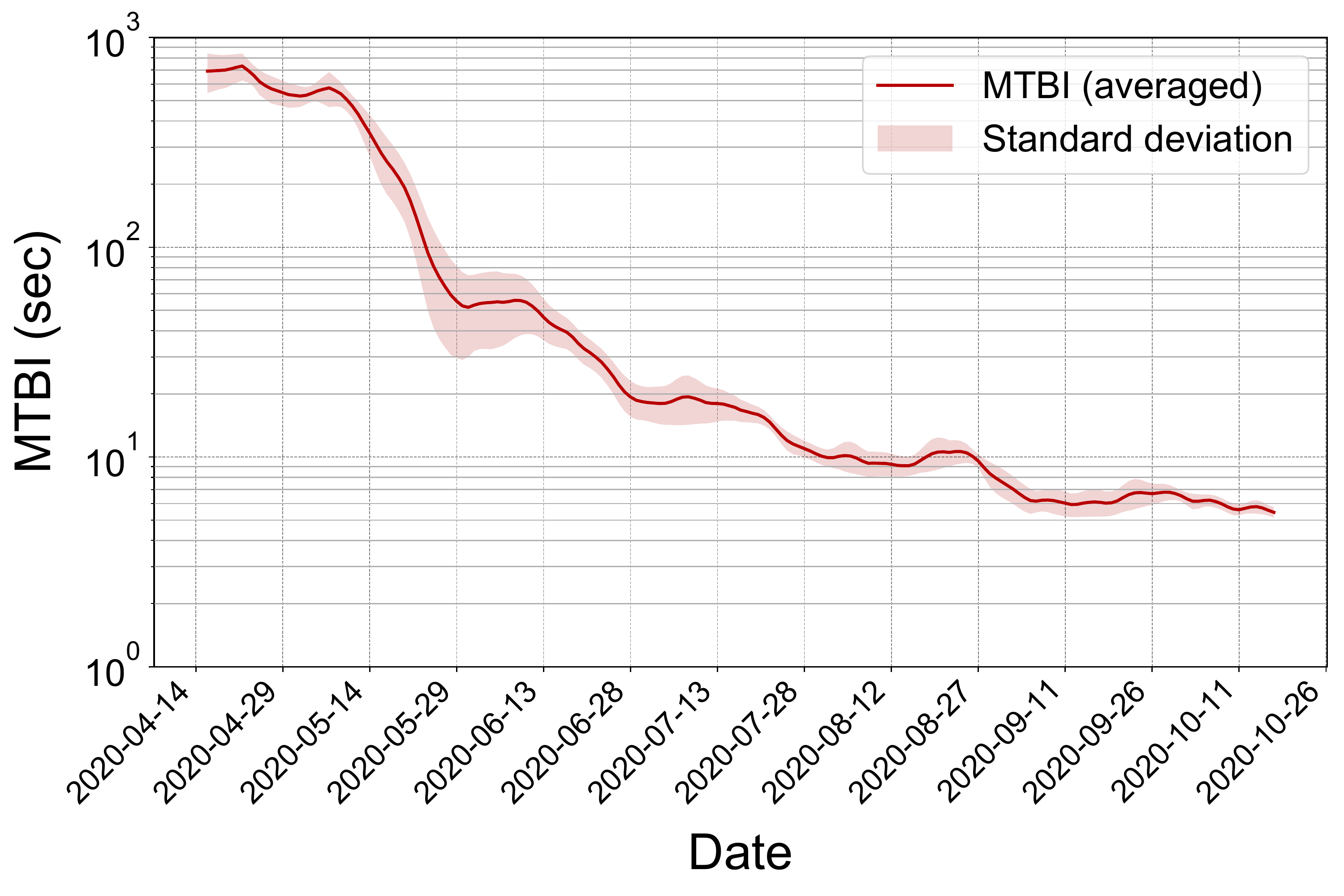}} 
		\subfloat[\centering Germany]{\label{ger_mtbi}\includegraphics[width = 0.33\linewidth]{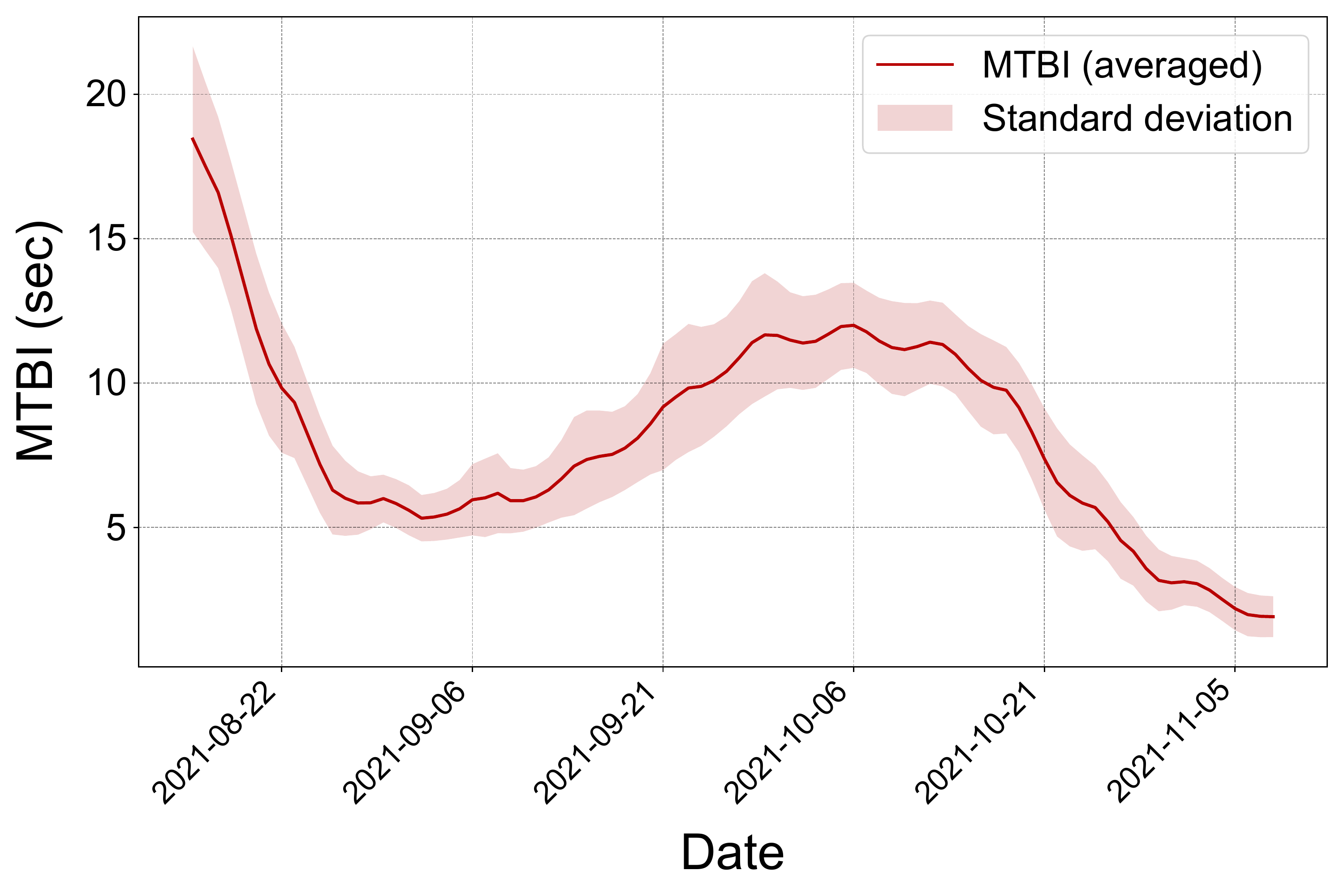}} 
		\subfloat[\centering United States]{\label{usa_mtbi}\includegraphics[width = 0.33\linewidth]{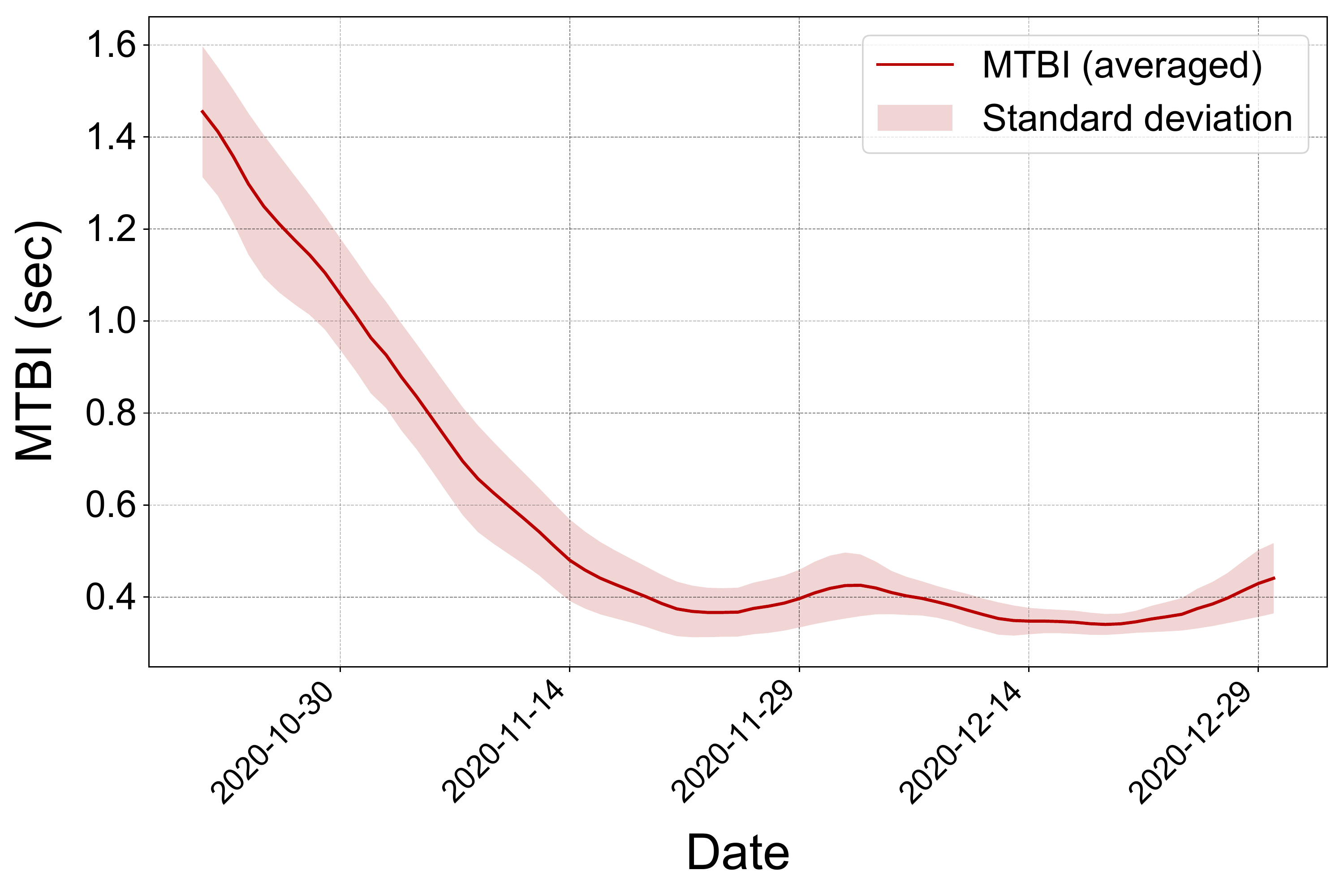}} 
		
		\caption{Evolution over time of the MTBI indicator for each of the three countries in its chosen stage. The solid curves were obtained by averaging 31 different estimations of the MTBI, one using the whole stage history and the rest considering time windows between 15 and 44 days. The light red shadow indicates the standard deviation within the 31 averaged curves. In the case of Argentina, the image is shown in a log scale to cover the high range of values.}
		\label{fig_mtbi}
	\end{figure}
	
	\begin{figure}[ht]
		\centering
		\subfloat[\centering Argentina $\rho$]{\label{arg_rho}\includegraphics[width = 0.33\linewidth]{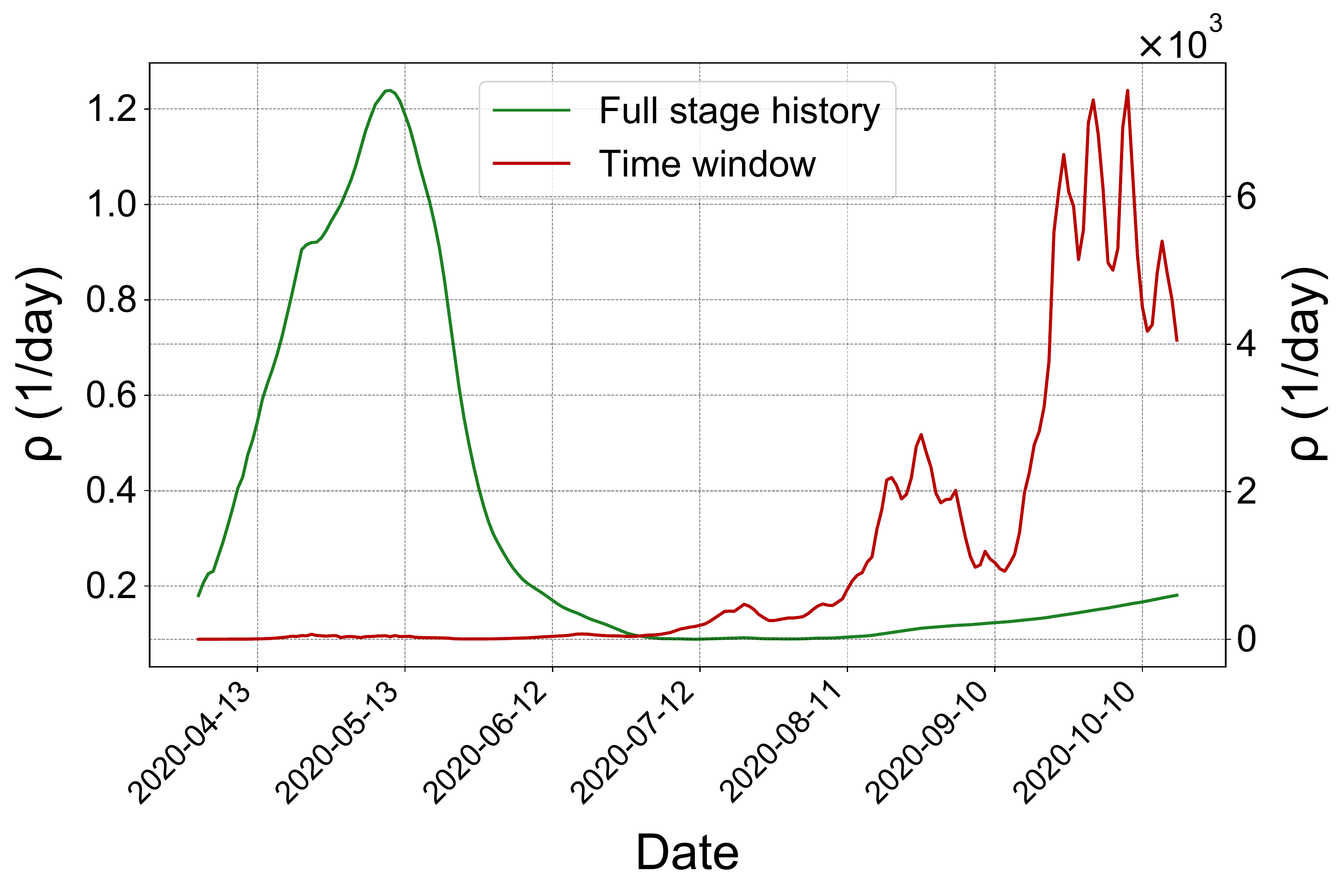}}
		\subfloat[\centering Germany $\rho$]{\label{ger_rho}\includegraphics[width = 0.33\linewidth]{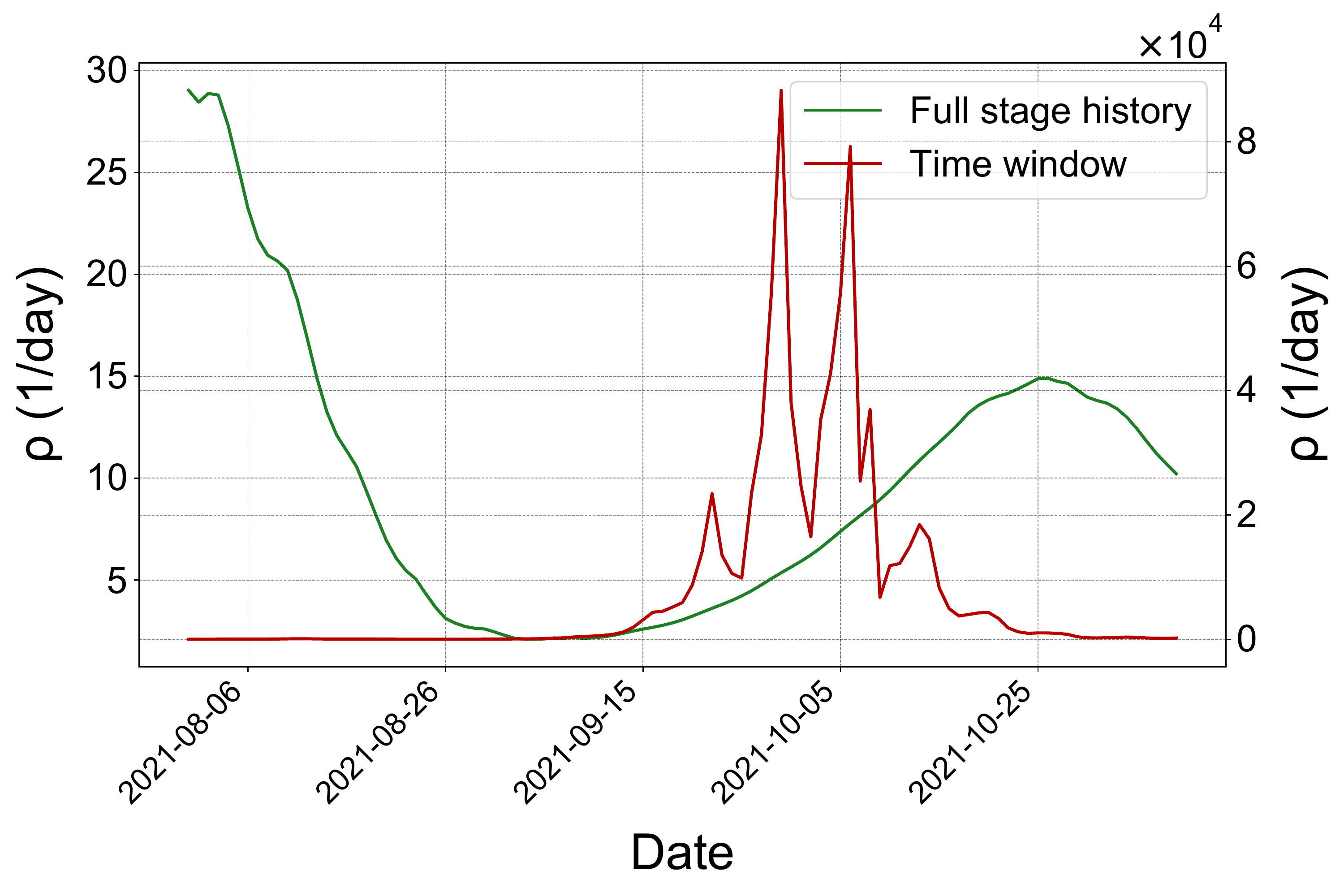}}
		\subfloat[\centering United States $\rho$]{\label{usa_rho}\includegraphics[width = 0.33\linewidth]{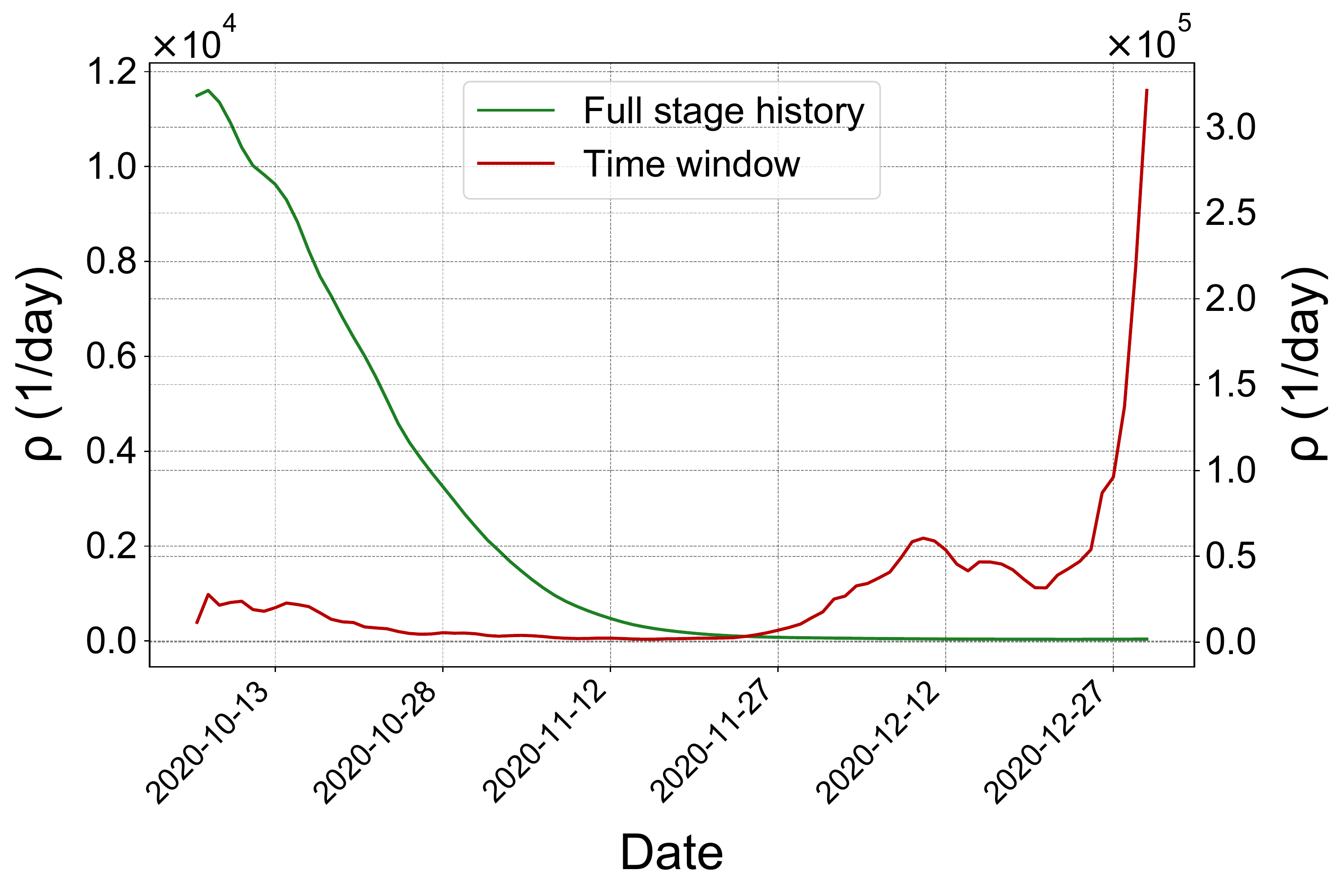}}
		\\
		\centering
		\subfloat[\centering Argentina $\gamma / \rho$]{\label{arg_gamma_per_rho}\includegraphics[width = 0.33\linewidth]{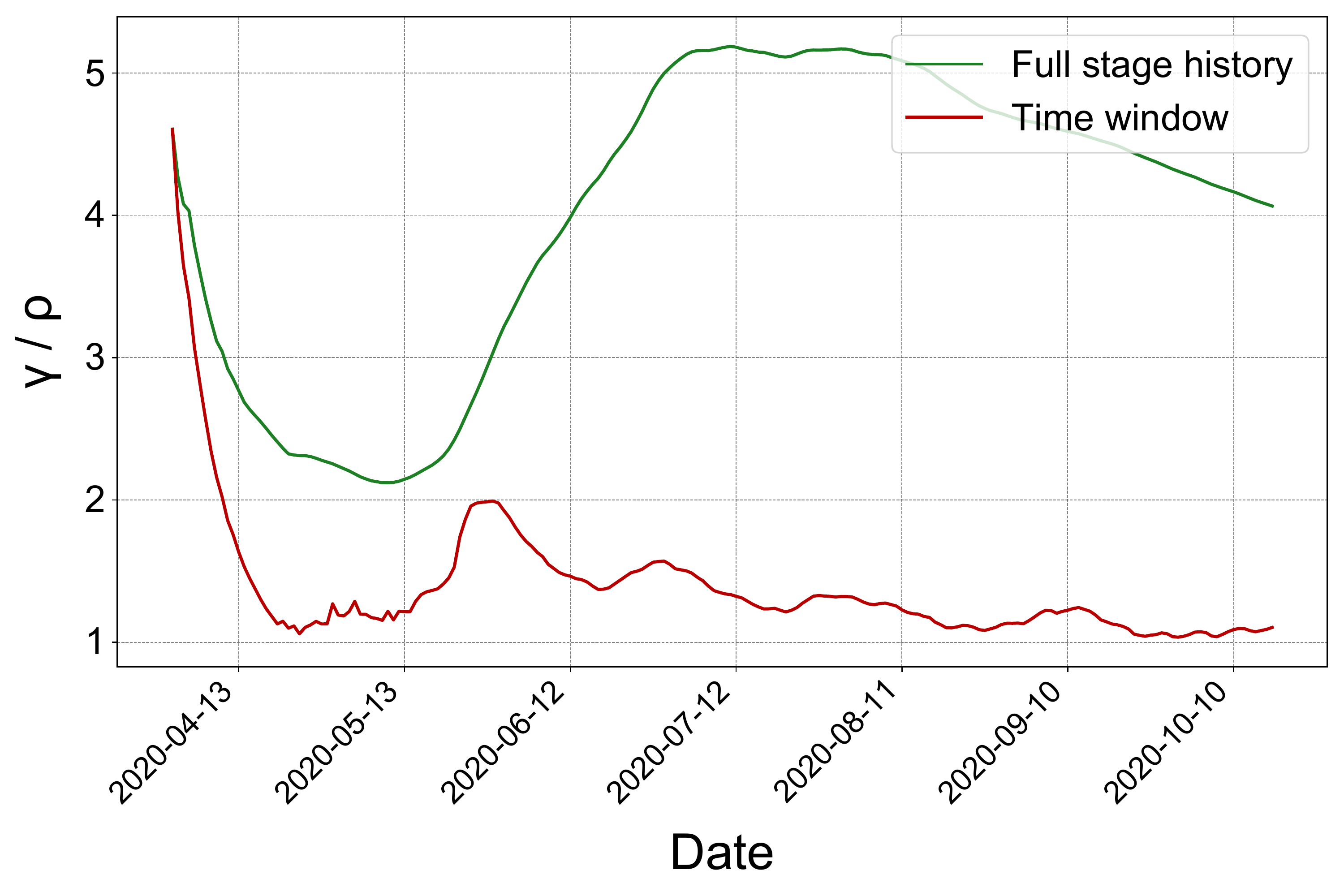}}
		\subfloat[\centering Germany $\gamma / \rho$]{\label{ger_gamma_per_rho}\includegraphics[width = 0.33\linewidth]{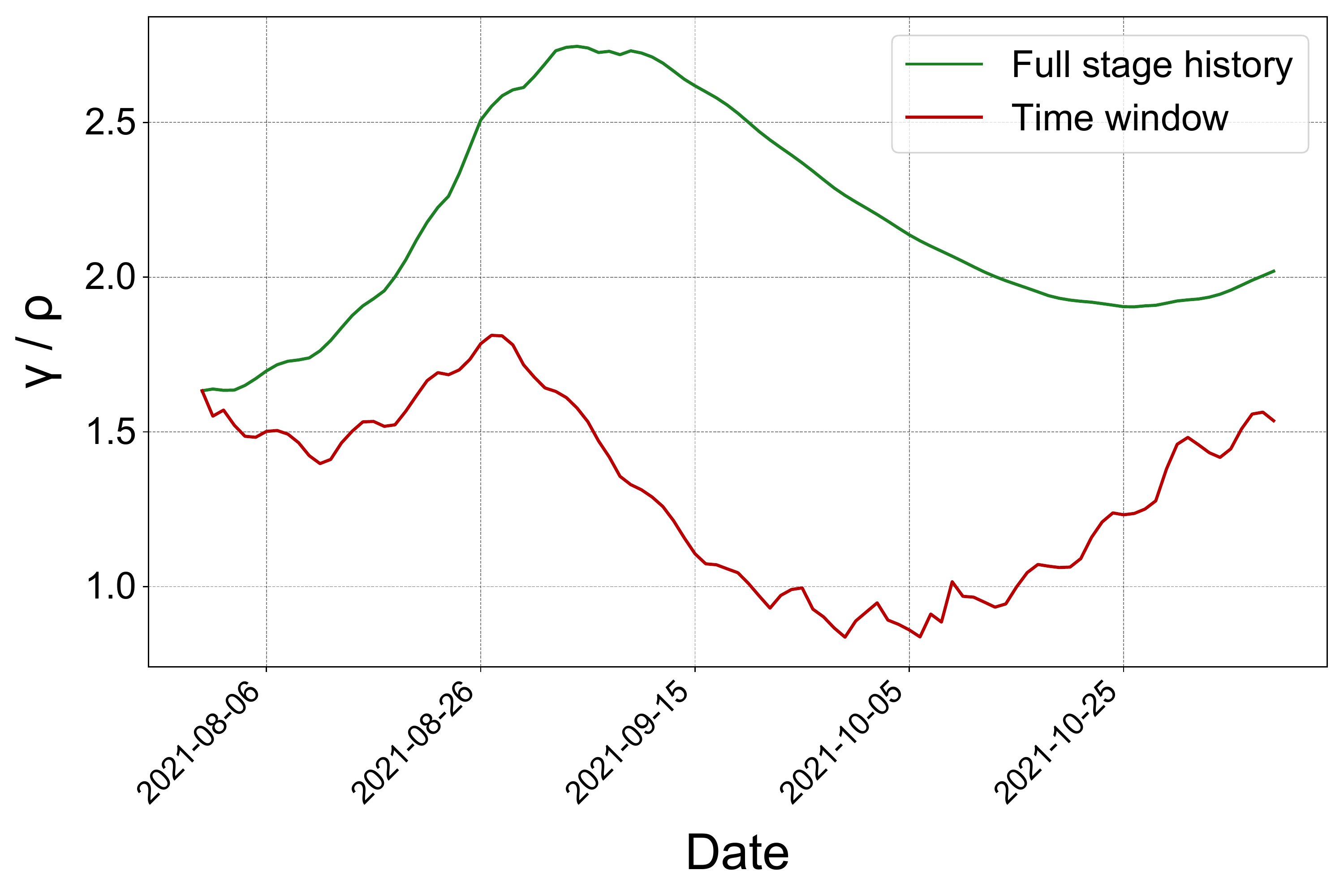}}
		\subfloat[\centering United States $\gamma / \rho$]{\label{usa_gamma_per_rho}\includegraphics[width = 0.33\linewidth]{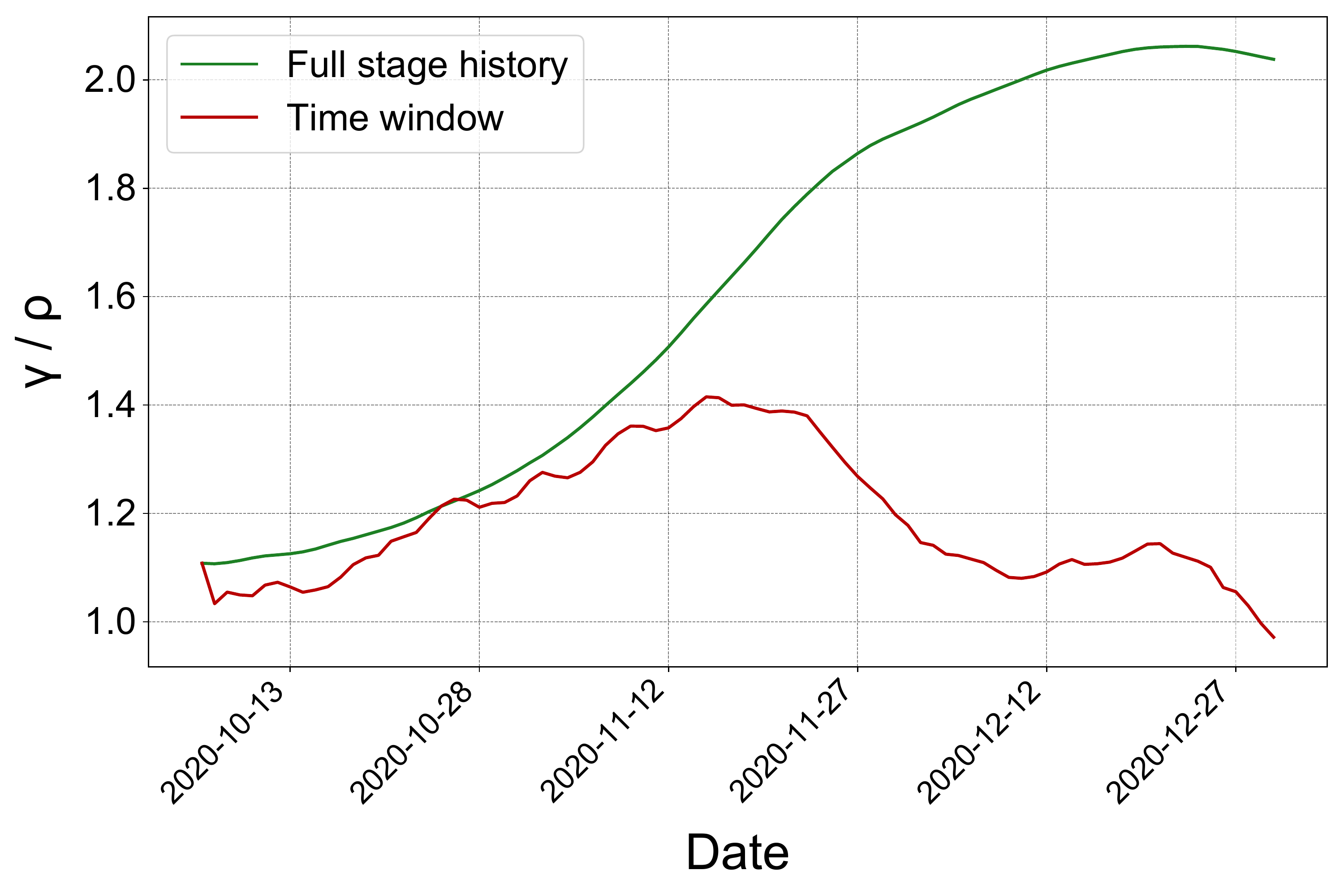}}
		
		\caption{Evolution over time of $\rho$ (upper row) and $\gamma/\rho$ (lower row) parameters for each of the three countries in its chosen stage. Green curves show the parameters calculated using the whole stage history. Red curves show the indicators calculated using a time window of 30 days. Since the $\rho$ numerical values are rather different on each case, the green curve scale is shown in the left side axis, whereas the red curve scale is shown in the right side axis for the three $\rho$ plots.} 
		\label{fig_parameters}
	\end{figure}
	
	\cref{fig_mtbi} shows the evolution over time of the averaged MTBI estimate and its variance for the chosen stage in each country. In all the study cases and with all the considered input datasets the model achieves an $R^2$ coefficient above $0.97$. For the three cases considered, only the corresponding stage for Germany has a vaccination process, rising from 35 to 50 \% over that period. The consequently sudden slow down in the speed propagation can be seen in the picture. Due to the population size and social mobility (between 20 and 30 \% under normal) in the considered stage of the United States, the propagation speed is much higher than in the other two study cases. In the case of Argentina, the strict lockdown imposed during the first wave of the epidemic is reflected on high MTBI values at the beginning of the first wave (i.e., a slow propagation). 
	
	It can also be seen in the three plots of \cref{fig_mtbi} that the MTBI curves do not change significantly in neither shape or scale when the data is subject to windowing, i.e., the MTBI indicator is invariant. This shows that the MTBI efficiently captures the behaviour of the model, and moreover it is robust to the length of the dataset used for calibration. Therefore, we propose to use the average of several estimations of the MTBI obtained through time windows of different sizes as our final MTBI estimator. 
	
	As it can be seen in \cref{fig_parameters}, the obtained $\rho$ and $\gamma/\rho$ parameters are indeed very different according to the estimation method used in each case. The resulting $\rho$ curve changes in both shape and scale (i.e., the numerical values have different orders of magnitude, which is why we plot each curve on its own $y$ axis), as it can be seen in \cref{arg_rho,usa_rho,ger_rho}. The $\gamma/\rho$ curve changes in shape, but the numerical values do not differ much (the pairs of curves in \cref{arg_gamma_per_rho,usa_gamma_per_rho,ger_gamma_per_rho} are plotted in the same scale); this implies that even if $\gamma$ and $\rho$ change individually between the two methods, they do it proportionally. This behaviour shows that computing the model parameters individually, either $\rho$ or $\gamma/\rho$, do not provide enough information about the epidemic, whereas the MTBI does.
	
	\section*{Discussion / Conclusions}
	
	In this work we have analyzed the evolution over time of the MTBI indicator given by a recently proposed stochastic non-homogeneous Markov model when applied to the COVID-19 pandemic, which is useful to measure the propagation speed of the epidemic. Three countries were considered as study cases: Argentina, Germany and the United States. For each of these countries, a particular stage was chosen and the model was fitted using different sets of input data. On the one hand, the full history of the stage until the day $t$ was considered to calibrate the model parameters at that day, and on the other hand, the calibration at $t$ was made using only the data from several time windows of different lengths. We show the evolution over time of the averaged MTBI indicator, as well as the evolution of the model parameters. The most important conclusion was that the MTBI indicator (which depends on those parameters) does not change significantly using different sizes of input data, which shows this is a robust indicator independently of the dataset size.

\bibliography{biblio}

\end{document}